\documentclass{aa}
\usepackage{graphicx}
\usepackage{txfonts}
\newcommand{\RSTAR}{\mbox{$R_{\star}$}}
\newcommand{\MSTAR}{\mbox{$M_{\star}$}}
\newcommand{\TEFF}{\mbox{$T_{\rm eff}$}}

\newcommand{\RSOL}{\mbox{$R_{\sun}$}}
\newcommand{\MSOL}{\mbox{$M_{\sun}$}}

\newcommand{\MSOLPERYR}{\mbox{$M_{\sun}$~yr$^{-1}$}}
\newcommand{\micron}{\mbox{$\mu$m}}
\newcommand{\KMS}{\mbox{km s$^{-1}$}}
\newcommand{\HOH}{\mbox{H$_2$O}}
\newcommand{\PERSQCM}{\mbox{cm$^{-2}$}}
\newcommand{\PERCBCM}{\mbox{cm$^{-3}$}}
\newcommand{\GRPERCBCM}{\mbox{g~cm$^{-3}$}}

\newcommand{\VMICRO}{\mbox{$\varv_{\rm micro}$}}

\newcommand{\LOGG}{\mbox{$\log \varg$}}
\newcommand{\ABUNDH}{\mbox{$\log A_{\rm H}$}}
\newcommand{\ABUNDC}{\mbox{$\log A_{\rm C}$}}
\newcommand{\ABUNDN}{\mbox{$\log A_{\rm N}$}}
\newcommand{\ABUNDO}{\mbox{$\log A_{\rm O}$}}
\begin{document}
\title{
Spatially resolved, high-spectral resolution observation 
of the \\
K giant Aldebaran in the CO first overtone lines
with VLTI/AMBER
\thanks{
Based on AMBER observations made with the Very Large Telescope 
Interferometer of the European Southern Observatory. 
Program ID: 090.D-0459(A)
}
}

\author{K.~Ohnaka
}

\offprints{K.~Ohnaka}

\institute{
Max-Planck-Institut f\"{u}r Radioastronomie, 
Auf dem H\"{u}gel 69, 53121 Bonn, Germany\\
\email{kohnaka@mpifr.de}
}

\date{Received / Accepted }

\abstract
{}
{
We present a high-spatial and high-spectral resolution observation 
of the well-studied K giant Aldebaran with AMBER at the Very Large Telescope 
Interferometer (VLTI).  
Our aim is to spatially resolve the outer atmosphere (so-called MOLsphere) 
in individual CO first overtone lines and derive its physical properties, 
which are important for understanding the mass-loss mechanism in normal 
(i.e., non-Mira) K--M giants.  
}
{
Aldebaran was observed between 2.28 and 2.31~\micron\ 
with a projected baseline length of 10.4~m and a spectral resolution of 12000.  
}
{
The uniform-disk diameter observed in the CO first overtone lines is  
20--35\% larger than is measured in the continuum.  
We have also detected a signature of inhomogeneities in the CO-line-forming 
region on a spatial scale of $\sim$45~mas, which is more than twice as 
large as the angular diameter of the star itself.  
While the MARCS photospheric model reproduces the observed spectrum well, 
the angular size in the CO lines predicted by the MARCS model is 
significantly smaller than observed.  This is because the MARCS model 
with the parameters of Aldebaran has a geometrical extension of 
only $\sim$2\% (with respect to the stellar radius).  
The observed spectrum and interferometric data in the CO lines can be 
simultaneously reproduced by placing an additional CO layer above the 
MARCS photosphere.  
This CO layer is extended to $2.5 \pm 0.3$~\RSTAR\ with CO column densities 
of $5\times10^{19}$--$2\times10^{20}$~\PERSQCM\ and a temperature of 
$1500\pm200$~K.  
}
{
The high spectral resolution of AMBER has enabled us to spatially resolve 
the inhomogeneous, extended outer atmosphere (MOLsphere) in the individual 
CO lines for the first time in a K giant.  
Our modeling of the MOLsphere of Aldebaran suggests a rather small 
gradient in the temperature distribution above the photosphere 
up to 2--3~\RSTAR.  
}


\titlerunning{Spatially resolving 
the K giant Aldebaran in the CO first overtone lines
}
\authorrunning{K. Ohnaka}
\maketitle

\begin{table*}
\begin{center}
\caption {Summary of the AMBER observation of Aldebaran with the B2-C1-D0 
  AT configuration.  The seeing and coherence time ($\tau_0$) are in the 
  visible. 
}
\vspace*{-2mm}

\begin{tabular}{l c c c c r c c c}\hline
Night  & $t_{\rm obs}$ & $B_{\rm p}$ & PA     & Seeing           & $\tau_0$ & 
Airmass & DIT & Number     \\ 
       & (UTC)        & (m)        & (\degr)& (\arcsec)        &  (ms)    &
        & (ms) & of frames \\
\hline
\multicolumn{9}{l}{Aldebaran}\\
\hline
2012 Dec 12 & 05:18:32 & 31.1/10.4/20.7 & 37/37/37 & 0.64 & 4.4 & 1.45 & 121 & 500 \\
\hline
\multicolumn{9}{l}{Procyon (calibrator)}\\
\hline
2012 Dec 12 & 05:40:16 & 28.1/9.4/18.7 & 22/22/22 & 0.98 & 2.9 & 1.21 & 121 & 2500 \\
\hline
\label{obslog}
\vspace*{-7mm}

\end{tabular}
\end{center}
\end{table*}

\begin{figure*}
\sidecaption
\rotatebox{0}{\includegraphics[width=12cm]{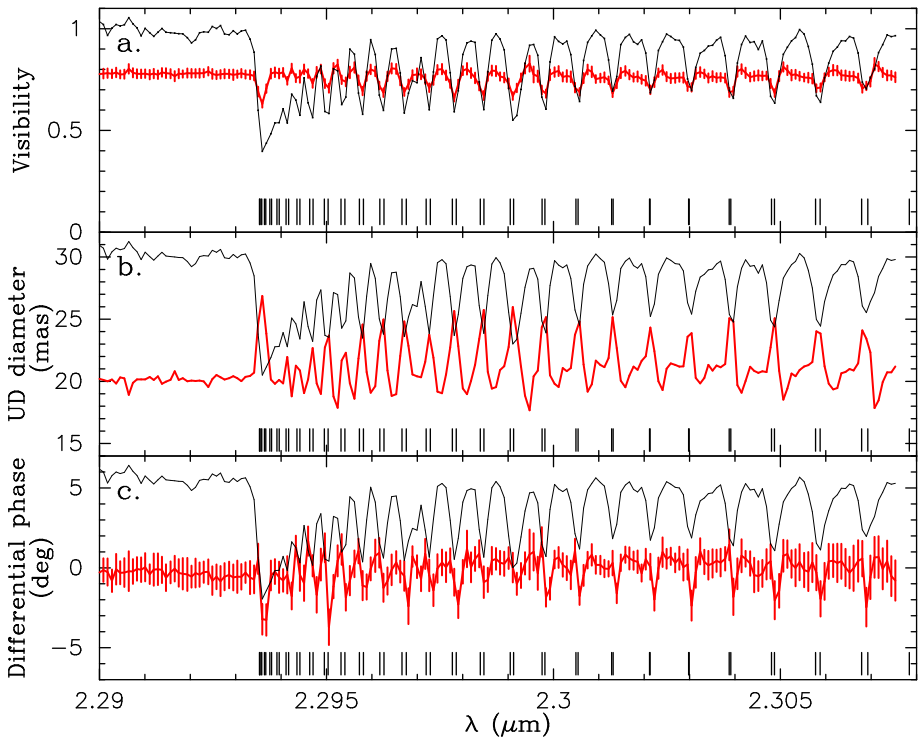}}
\caption{
AMBER observation of Aldebaran in the CO first overtone lines. 
{\bf a:} Visibility observed on the 10.4~m baseline (red solid line). 
{\bf b:} Uniform-disk (UD) diameter (red solid line). 
{\bf c:} Differential phase (red solid line).  
In each panel, the scaled observed spectrum is shown by the black solid line. 
The positions of the CO lines are also marked with the ticks. 
}
\label{obsres}
\end{figure*}

\section{Introduction}
\label{sect_intro}

Red giant stars, which represent the late evolutionary stages of 
intermediate- to low-mass stars from the red giant branch (RGB) to the
asymptotic giant branch (AGB), experience mass loss ranging from 
$10^{-11}$ \MSOLPERYR\ up to $10^{-4}$ \MSOLPERYR.  
However, the mass loss in red giants is not yet understood well.  
For Mira-type AGB stars with large variability  amplitudes 
($\Delta V \approx 9$), the combination of the levitation of the atmosphere 
by the stellar pulsation and the radiation pressure on dust grains is 
often believed to drive the mass loss (see, e.g., H\"ofner \cite{hoefner11} 
for discussion of the success and the current problem with this mechanism).  
However, it is by no means clear whether this mechanism can operate in 
red giants in general, that is, normal (i.e., non-Mira-type) K and M giants, 
whose variability amplitudes ($\Delta V \approx 1$--2) are significantly 
smaller than Miras.  Furthermore, K giants and early M giants are experiencing 
mass loss, despite the absence of dust.  

Observations of the outer atmosphere are important for understanding 
the mass-loss mechanism.  
Infrared spectroscopic studies of molecular lines reveal the presence 
of an extended molecular outer atmosphere, 
the so-called MOLsphere, in K and M giants, as well as in red supergiants 
(Tsuji \cite{tsuji88}, \cite{tsuji00a}, \cite{tsuji00b}, \cite{tsuji01}; 
Tsuji et al. \cite{tsuji97}).  
%
Spectro-interferometry, which combines high spectral resolution 
and high spatial resolution, is powerful for studying the physical 
properties of the MOLsphere.  

We spatially resolved the normal M7 giant BK Vir in the 
CO first overtone lines near 2.3~\micron\ 
with a spectral resolution of 12000 using VLTI/AMBER 
(Ohnaka et al. \cite{ohnaka12}).  
The observed CO line spectrum can be reproduced well by the MARCS 
photospheric models (Gustafsson et al. \cite{gustafsson08}), which are 
state-of-the-art models for the photosphere of cool evolved stars.  
However, the MARCS models clearly fail to explain the observed angular 
size of the star in the individual CO lines. 
The observed angular size in the CO lines is much greater than the models 
predict, suggesting the presence of more extended components. 
We found out that the observed spectrum and angular size can be explained 
by a model in which 
two extra CO layers are added above the MARCS model photospheres, 
at 1.2--1.25~\RSTAR\ and 2.5--3~\RSTAR\ with CO column 
densities of $10^{22}$~\PERSQCM\ and $10^{19}$--$10^{20}$~\PERSQCM.  
Surprisingly, the temperatures of the CO layers (1900--2100~K and 
1500--2100~K) are higher than or equal to the temperature of the uppermost 
layer of the 
photosphere (see Fig.~9 of Ohnaka et al. \cite{ohnaka12}).  
While many K and M giants possess chromospheres, the derived temperatures 
of the MOLsphere are still lower than in the chromosphere 
($\sim$6000--10000~K).  Therefore, although the MOLsphere shows a 
temperature inversion, it is still different from the chromosphere.  

A better understanding of the physical properties of the MOLsphere 
is important for clarifying the mass-loss mechanism, because the 
physical process responsible for forming the MOLsphere may 
be related to the driving mechanism of the mass loss.  
Given that IR spectro-interferometric observations of normal K and 
M giants are still scarce, 
AMBER observations of more normal red giants are indispensable for 
shedding new light on the origin of the MOLsphere.  

As a first step, we present a high-spatial and high-spectral resolution 
observation of the CO first overtone lines in 
the well-studied K5 giant Aldebaran  ($\alpha$~Tau) with 
VLTI/AMBER.  
%
%
At a distance of 20.4~pc (parallax = 48.94~mas, van Leeuwen 
\cite{vanleeuwen07}), Aldebaran is one of the nearby red giants.  
Its angular diameter has been measured by the lunar occultation 
and long-baseline optical/IR interferometry. 
As summarized in Richichi \& Roccatagliata (\cite{richichi05}), 
most of the angular diameter measurements between the $V$ and 
$L$ band indicate 18--22~mas but without a noticeable wavelength 
dependence.  As discussed in their paper, 
the scatter in the measured angular diameters mostly reflects 
the discrepancies among the lunar occultation measurements, and 
the scatter among the diameters measured by long-baseline 
interferometry is smaller.  
In the present work, we adopt the uniform-disk diameter of 
$19.96\pm0.03$~mas derived by 
Richichi \& Roccatagliata (\cite{richichi05}), which is the average 
of their lunar occultation measurements at 2.22 and 3.55~\micron\ 
and the VLTI/VINCI measurements in the $K$ band.  
They argue that their measurements would represent the best available 
value for the angular diameter of Aldebaran.  
We adopt an effective temperature (\TEFF) of $3874\pm100$~K and a stellar 
mass (\MSTAR) = $1.5\pm0.3$~\MSOL\ from Tsuji (\cite{tsuji08}) for the 
present work.  Combining this stellar mass and the radius 
of 44~\RSOL\ measured by Richichi \& Roccatagliata (\cite{richichi05}) 
results in a surface gravity of \LOGG\ = 1.5 (in units of cm~s$^{-2}$).  
The metallicity of Aldebaran is solar or marginally subsolar with [Fe/H] = 
$-0.14\pm0.30$ (Kov\'acs \cite{kovacs83}) and 
$-0.15\pm0.2$ (Decin et al. \cite{decin03}).  
Based on high-resolution spectra in the $H$, $K$, and $L$ bands, 
Tsuji (\cite{tsuji08}) derived the carbon, nitrogen, and oxygen abundances 
to be \ABUNDC\ = $8.38\pm0.04$, \ABUNDN\ = 8.05, and \ABUNDO\ = 
$8.79\pm0.04$ (abundances are expressed on the logarithmic scale 
with \ABUNDH\ = 12).  
These CNO abundances suggest the mixing of the CN-cycled 
material in the first dredge-up, although the derived $^{12}$C/$^{13}$C 
ratio of 10 cannot be fully understood by the standard evolutionary theory. 
The micro-turbulent velocity derived in previous studies 
is approximately 2~\KMS\ (Smith \& Lambert \cite{smith85}; 
Decin et al. \cite{decin03}; Tsuji \cite{tsuji08}).  
The mass-loss rate and the terminal wind velocity of Aldebaran are estimated 
to be (1--1.6)$\times 10^{-11}$~\MSOLPERYR\ 
(Robinson et al. \cite{robinson98}; Wood et al. \cite{wood07}) and 
30~\KMS\ (Robinson et al. \cite{robinson98}), respectively.  

Tsuji (\cite{tsuji01}) discovered the \HOH\ absorption feature at 
$\sim$6.6~\micron\ in Aldebaran, which originates in the MOLsphere.  
Based on a detailed analysis of a high-resolution spectrum of Aldebaran, 
Tsuji (\cite{tsuji08}) revealed absorption excess in the strong CO 
first overtone lines, which is also interpreted as the contribution 
of the MOLsphere.  
On the other hand, the emission lines in the UV (e.g., \ion{Mg}{ii}, 
\ion{Fe}{ii}, and \ion{O}{i} lines) indicate the presence of 
the chromosphere.  
This inhomogeneous, multi-component nature of the outer atmosphere is 
consistent with what is suggested from 
the observations of the CO fundamental lines in the IR 
(Wiedemann et al. \cite{wiedemann94}) and the CO fluorescent lines in the 
UV (McMurry \& Jordan \cite{mcmurry00}).  
Note, however, that the nature of the CO gas probed by the 4.6~\micron\ 
fundamental lines ($\varv = 1 -0$) 
might be different from that probed by the first overtone lines 
($\varv = 2 -0$) near 2.3~\micron\ in the present work.    
The spectral energy distribution (SED) of Aldebaran from the optical to 
$\sim$1~mm, as well as the mid-IR spectrum 
shows no signature of dust (Dehaes et al. \cite{dehaes11}; Monnier et al. 
\cite{monnier98}), and therefore, this star is ideal for studying the 
mass-loss mechanism without dust.


\section{Observation}
\label{sect_obs}

AMBER is a spectro-interferometric instrument at VLTI and can provide 
high spatial resolutions and high spectral resolutions of up to 12000 
(Petrov et al. \cite{petrov07}) by combining three 8.2~m Unit Telescopes 
(UTs) or 1.8~m Auxiliary Telescopes (ATs).  
AMBER measures the visibility, which is the amplitude of the Fourier 
transform of the object's intensity distribution on the sky, as well as 
the closure phase and differential phase.  
The closure phase contains information about the asymmetry of the object, with 
non-zero and non-$\pi$ closure phases being the signature of asymmetry. 
The differential phase provides information about the photocenter shift of
the object in spectral features with respect to the continuum.  

We observed Aldebaran on 2012 December 12 with AMBER using the B2-C1-D0 
(11--23--34m) linear array (Program ID: 090.D-0459A, P.I.: K.~Ohnaka), 
as summarized in Table~\ref{obslog}.  
We used the high-spectral resolution mode in the $K$ band (HR\_K) with 
a spectral resolution of 12000 between 2.28 and 2.31~\micron\ to 
observe the CO first overtone lines near the 2--0 band head at 
2.294~\micron.  The VLTI fringe tracker FINITO was not used, 
because while the visibility in the $H$ band, where FINITO operates, 
is higher than the limit of 0.15 on the shortest baseline, 
the $H$-band visibility on the second shortest baseline is below the limit. 
However, the high brightness of Aldebaran 
enabled us to detect fringes without FINITO with a Detector Integration 
Time (DIT) of 121~ms.  
We observed Procyon (F5IV-V) not only as the interferometric calibrator 
but also as the spectroscopic calibrator.  

We reduced the data using amdlib ver.3.0.5\footnote{
Available at http://www.jmmc.fr/data\_processing\_amber.htm} as 
described in Ohnaka et al. (\cite{ohnaka09}). 
We took the best 80\% of the frames in terms of the fringe S/N 
(Tatulli et al. \cite{tatulli07}).  
For the calibrator Procyon, we adopted the uniform-disk diameter of 
$5.38 \pm 0.05$~mas from the CHARM2 catalog (Richichi et al. \cite{charm2}).  
There are two issues that had to be taken into account in the 
reduction of the present data.  
Firstly, 
the telescope at the D0 station was shadowed by UT1 during the observation 
of Aldebaran (but not during the observation of Procyon).  
Therefore, the flux of Aldebaran at this telescope was lower than at the 
other two telescopes by a factor of $\sim$10, which makes the derivation of 
the interferometric observables on the baselines including the D0 telescope 
less reliable.  Therefore, in the present work, we only discuss the 
visibility and differential phase obtained on the B2--C1 baseline, which 
provides a projected baseline length ($B_{\rm p}$) of 10.4~m and a position 
angle (PA) of 37\degr. 
Secondly, 
only one data set of the calibrator was obtained, which makes it 
difficult to carry out a reliable calibration of the absolute visibility 
level.  
Because the time variation expected in the angular diameter of Aldebaran 
is as small as $\pm0.17$~mas (Richichi \& Roccatagliata \cite{richichi05}), 
we scaled the observed calibrated visibility in 
the continuum to those expected from a uniform disk with the adopted angular 
diameter of 19.96~mas.  

The wavelength calibration was carried out using the telluric lines 
identified in the observed spectrum of Procyon, as described in 
Ohnaka et al. (\cite{ohnaka12}).  The uncertainty in the wavelength 
calibration is $3.2\times10^{-5}$~\micron\ (4.2~\KMS).  
The wavelength scale was converted 
to the laboratory frame using a heliocentric velocity of 54.2~\KMS\ 
of Aldebaran (Gontcharov \cite{gontcharov06}).  
The spectrum of Procyon in the observed spectral window is featureless 
with AMBER's spectral resolution except for the Mg line at 2.2814~\micron.  
Therefore, the calibrated 
spectrum of Aldebaran was obtained by dividing the observed spectrum 
of Aldebaran with that of Procyon.

%

\section{Results}
\label{sect_res}


Figure~\ref{obsres}a shows the observed visibility in the CO lines on the 
10.4~m baseline with the observed (spatially unresolved) spectrum.  
The signature of the CO lines is clear in the observed 
visibility.  The uniform-disk diameter derived from the visibility 
is shown in Fig.~\ref{obsres}b.  The uniform-disk diameter in the CO lines 
is 20--35\% larger than in the continuum.   
Figure~\ref{obsres}c shows the observed differential phase.  We detect 
non-zero differential phases in the CO lines (i.e., wavelength-dependent 
photocenter shifts), which indicate the presence of 
asymmetric or inhomogeneous structures in the CO-line-forming region 
on a spatial scale of $\sim$45~mas (corresponding to the spatial 
resolution with the 10.4~m baseline).  
This is more than twice as large as the angular diameter of the 
star itself and consistent with the size of the MOLsphere as discussed 
in Sect.~\ref{sect_modeling}. 
Richichi \& Roccatagliata (\cite{richichi05}) found possible surface 
structures---though not conclusive as the authors mention---in the 1-D 
image of Aldebaran reconstructed from the lunar 
occultation data taken at 2.22~\micron\ with a spectral resolution of 
$\sim$70, which sample the spectral region free from strong 
lines.  
Our AMBER observation is the first study to spatially resolve 
the atmosphere of a K giant in individual CO lines and detect asymmetry 
in the CO-line-forming region.  

The observed increase in the uniform-disk diameter in the CO lines 
is in marked contrast to the negative detection of a change in the 
angular size in the TiO band at 7120~\AA\ (Quirrenbach et al. 
\cite{quirrenbach93}).  However, this might be due to the lower spectral 
resolution used in their observations.  
For example, if we bin our AMBER data to a spectral resolution of 1500 
as described in Ohnaka et al. (\cite{ohnaka09}), the increase in the 
uniform-disk diameter in the CO band head is 4\%.  
The narrow-band filter for the TiO band used by Quirrenbach et al. 
(\cite{quirrenbach93}) has a spectral resolution of $\sim$60.  
The width of the filter (120~\AA) is smaller than the width of the 
TiO band ($\sim$400~\AA) starting at $\sim$7050~\AA.  However, 
the observed spectrum of Aldebaran in Kieling (\cite{kiehling87}, 
Fig.~4c, HR~1457 = Aldebaran) shows that this TiO band is centered 
at 7200~\AA, and the central wavelength of the filter (7120~\AA) does not 
exactly match the center of the TiO band.  
The filter samples the spectral region between the 
beginning and the center of the TiO band, not just the deepest 
part of the TiO band.  
Therefore, the low spectral resolution, as well as the shift 
between the filter's central wavelength and the center of the TiO 
band, might have led to the negative detection of 
the extended atmosphere of Aldebaran.  
It might not have been a problem for 
detecting a change in the angular size for cooler stars with more 
pronounced extended atmospheres. 
In any case, measurements of the angular diameter of K giants in the 
TiO bands with higher spectral resolutions 
are necessary to compare with the angular diameter in the 
CO first overtone lines.  
This can be feasible, for example, with the visible interferometric 
instrument VEGA 
at CHARA with spectral resolutions of up to 30000 (Mourard et al. 
\cite{mourard09}).

The high spatial and high spectral resolution of AMBER 
has enabled Ohnaka et al. (\cite{ohnaka09}, \cite{ohnaka11}, and 
\cite{ohnaka13a}) to spatially resolve the gas motions in the photosphere 
and MOLsphere in the red supergiants Betelgeuse and Antares.  
However it is difficult to carry out a similar analysis for the current 
data of Aldebaran, 
because the amplitude of the velocity field in the photosphere (and 
possibly also in the MOLsphere) is much smaller than in the 
red supergiants.  For example, Gray (\cite{gray09}) measured a time variation 
in the radial velocity of $\pm$0.3~\KMS\ in Aldebaran, which is much 
smaller than the amplitude of $\sim$7~\KMS\ found in Betelgeuse 
(Gray \cite{gray08}).  
The micro- and macro-turbulent velocities of $\sim$2~\KMS\ and 
$3.6\pm0.3$~\KMS\ derived for Aldebaran (references given in Sect.~\ref{sect_intro} 
and Tsuji \cite{tsuji86}) are also 
noticeably smaller than the micro- and macro-turbulent velocities of $\sim$5 
and 10--20~\KMS\ in Betelgeuse (Ohnaka et al. \cite{ohnaka09} and references 
therein).  
The amplitude of the velocity field suggested from these observations 
is too small compared to AMBER's spectral resolution and the accuracy of 
the wavelength calibration.

\begin{figure}
\resizebox{\hsize}{!}{\rotatebox{0}{\includegraphics{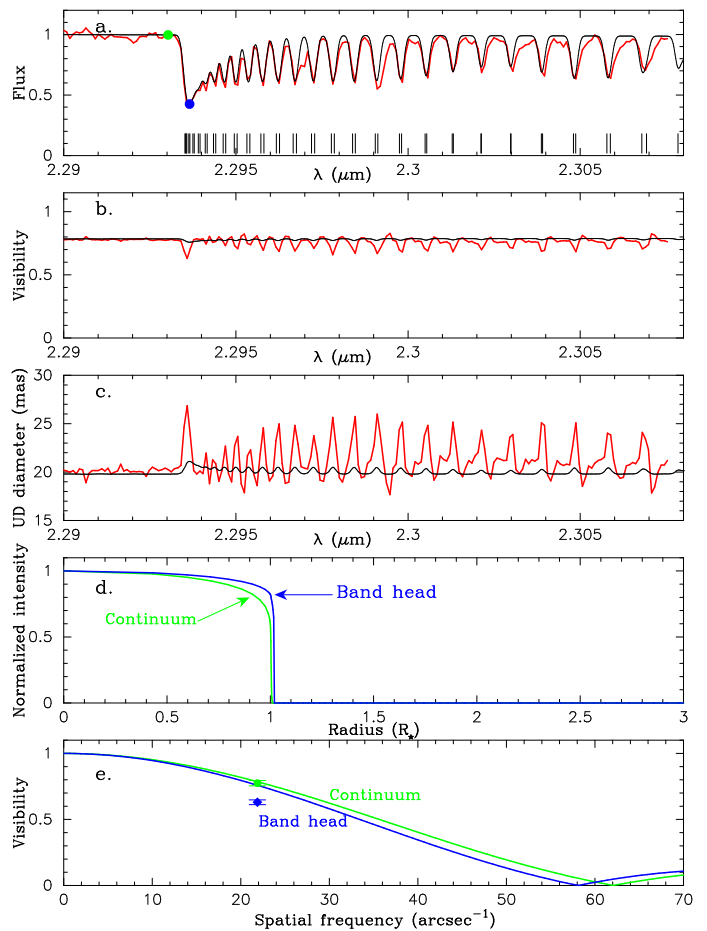}}}
\caption{
Comparison of the MARCS model with the AMBER data of Aldebaran. 
The parameters of the MARCS model are given in Sect.~\ref{sect_modeling}.  
{\bf a:} Model spectrum (black solid line) 
computed from the pressure and temperature distributions of the 
MARCS model as described in Appendix \ref{appendix_model} 
(not the pre-computed spectrum available on the MARCS website)  
with the observed spectrum 
(red solid line).  The green and blue dots show the wavelengths of the 
intensity profiles and visibilities shown in the panels {\bf d} and {\bf e}.  
The positions of the CO lines are marked by the ticks. 
{\bf b:} Model visibility predicted at the 10.4~m baseline (black solid 
line) with the observed data (red solid line).  
{\bf c:} Uniform-disk (UD) diameter derived from the MARCS model visibility 
shown in the panel {\bf b} (black solid line) with the one observed (red solid 
line).  
{\bf d:} Model intensity profiles predicted in the continuum (green solid 
line) and in the CO band head (blue solid line), at the wavelengths marked 
in the panel {\bf a}.  
{\bf e:} Model visibilities as a function of spatial frequency predicted in the 
continuum (green solid line) and in the CO band head (blue solid line).  
The visibilities 
observed at these wavelengths are plotted by the dot (continuum) and 
the filled diamond (CO band head).  
}
\label{marcs_model}
\end{figure}

\section{Modeling the AMBER data}
\label{sect_modeling}


We first compare the observed spectrum and visibility with the 
MARCS photospheric model (Gustafsson et al. \cite{gustafsson08}).  
The MARCS models represent plane-parallel or spherical photospheres 
in hydrostatic 
and radiative equilibrium, with the effects of molecular and atomic 
line opacities included using the opacity sampling method.  
Each MARCS model is specified by effective temperature (\TEFF), 
surface gravity (\LOGG), micro-turbulent velocity (\VMICRO), 
chemical composition, and (for the spherical case) 
stellar mass (\MSTAR).  
We selected spherical MARCS models with \TEFF\ = 3900~K, \LOGG\ = 1.5, 
\MSTAR\ = 1 and 2~\MSOL\ (no MARCS model with 1.5~\MSTAR\ is available), 
\VMICRO\ = 2~\KMS, 
and the ``moderately CN-cycled'' composition with [Fe/H] = 0.0, 
because these models have the parameters closest to 
those of Aldebaran summarized in Sect.~\ref{sect_intro}.  

Using the temperature and pressure distributions from the MARCS models, 
we computed the monochromatic intensity profile and visibility, as well 
as the spectrum using the CO line list of Goorvitch (\cite{goorvitch94}), 
as described in Appendix~\ref{appendix_model}.  
The angular scale of the models was 
determined so that the uniform-disk diameter derived from the 
model visibility in the continuum agrees with the adopted value of 
19.96~mas.  

Comparison of the MARCS model visibility and spectrum with the AMBER 
observation is shown in Fig.~\ref{marcs_model}.  
The observed CO line spectrum (Fig.~\ref{marcs_model}a) is 
reproduced well by the MARCS model 
with \TEFF\ = 3900~K, \LOGG\ = 1.5, and \MSTAR\ = 1~\MSOL.  
However, as Figs.~\ref{marcs_model}b and \ref{marcs_model}c show, 
the model predicts only marginal decrease in the visibility 
(thus only marginal increase in the uniform-disk diameter) 
in the CO lines and fails to explain the observed data.  
This means that the star is much more extended than the MARCS model 
predicts.  
The geometrical extension of the MARCS model, which is defined as 
the geometrical distance between the layers with $\tau_{\rm Ross} = 1$ 
and $\tau_{\rm Ross} = 10^{-5}$ ($\tau_{\rm Ross}$ is the optical depth 
based on the Rosseland mean opacity), is 2\% of the stellar radius.  
This geometrically thin photosphere is seen in Fig.~\ref{marcs_model}d, 
where the intensity profiles in the continuum and in the CO band head 
are shown.  
This means that the visibilities predicted in the continuum and in the CO 
band head differ only slightly (Fig.~\ref{marcs_model}e).  
This is why the 
MARCS model cannot explain the increase in the uniform-disk diameter 
of 20--35\% in the CO lines.  
We also examined whether MARCS models with slightly different parameters 
can explain the observed visibility.  However, the MARCS models with 
\TEFF\ = 3800--4000~K, \LOGG\ = 1.0 and 1.5, and \MSTAR\ = 1 and 2~\MSOL\ 
are characterized by the geometrical extension of 4\% at most, and 
therefore, cannot explain the observed significant increase in the angular 
size in the CO lines.  
This finding that the MARCS model can reproduce 
the CO first overtone line spectrum well but fails to explain the angular 
size of the star in the CO lines is the same as what we found in 
the M7 giant BK~Vir in Ohnaka et al. (\cite{ohnaka12}).  
Our AMBER observation with high spectral resolution has clearly revealed 
the presence of the MOLsphere, which cannot be accounted for by 
the photospheric models.


We derived the physical properties of the MOLsphere using the model 
described in Ohnaka et al. (\cite{ohnaka12}).  
In this model, one or two CO layers are added above the MARCS photospheric 
model.  
Given that the visibility was obtained only on one baseline in the present 
work, we used only one layer to keep the number of free 
parameters as small as possible.  
The geometrical thickness of the additional CO layer is 
fixed to 0.1~\RSTAR, and the micro-turbulent velocity in the CO layer is 
set to 2~\KMS, which is the same as in the MARCS photospheric model.  
The (inner) radius, CO column density in the radial direction, 
and temperature of the CO layer were treated as free parameters. 
The computation of the intensity profile, visibility, and spectrum from 
the MARCS+MOLsphere model is described in Appendix~\ref{appendix_model}. 


Figure~\ref{wme_model} shows a comparison of the best-fit MARCS+MOLsphere 
model with the observed spectrum and visibility.  The MOLsphere of this 
model is characterized by a radius of 2.5~\RSTAR\ with a temperature of 
1500~K and a CO column density of $1\times10^{20}$~\PERSQCM.  
The observed spectrum and visibility, as well as the uniform-disk diameter 
are reasonably reproduced.  
The intensity profile shown in Fig.~\ref{wme_model}d reveals the extended 
emission from the MOLsphere. 
While the intensity of this extended emission is low compared to the 
intensity on the stellar disk, the flux contribution is not 
negligible because it is very extended.  
As a result, the MOLsphere makes the star appear much more extended than 
the photosphere alone and can explain the observed angular size in the 
CO lines.  
We computed MOLsphere models with radii between 1.2 and 2.8~\RSTAR, 
temperatures between 1000 and 2000~K, and CO column densities between 
$1\times10^{19}$ and $1\times10^{21}$~\PERSQCM, and 
estimate the uncertainties in the radius, temperature, 
and CO column density to be $\pm 0.3$~\RSTAR, $\pm 200$~K, 
and a factor of 2, respectively.

It may appear contradictory that the model spectrum is hardly affected 
by the MOLsphere, while the angular size is largely affected.  
This can be explained as follows.  
Figure~\ref{alftau_1dspec} shows spatially resolved spectra predicted 
at two positions (at the center of the stellar 
disk and at the edge of the MOLsphere) by the MARCS+MOLsphere model, 
together with the spatially unresolved model spectrum.  
On the one hand, 
the additional CO layer introduces stronger absorption (i.e., CO absorption 
lines are deeper) for lines of sight within the stellar disk.  
This can be seen in the spatially resolved spectrum at the stellar disk 
center (red line) 
when compared to the spatially unresolved spectrum from the MARCS-only 
model (black line).  
On the other hand, 
the spatially resolved spectrum at the edge of the MOLsphere (blue line) 
shows that the CO lines appear in emission off the limb of the star 
(note that the off-limb spectrum is scaled up by a factor of 10 
for visual clarity).  
This extended emission from the MOLsphere fills in the additional absorption 
due to the MOLsphere itself, which results in little change in the 
spatially unresolved spectrum.  
The spectra predicted by the MARCS-only and MARCS+MOLsphere 
models shown in Figs.~\ref{marcs_model}a and \ref{wme_model}a show very 
little difference.  However, the difference in the visibility predicted 
by these models is obvious.  
This illustrates that the MOLsphere can become invisible for a spectrometer 
but not for a spectro-interferometer 
(however, signatures of the MOLsphere can be detected by spectroscopy 
for some lines, as demonstrated by Tsuji \cite{tsuji01}, \cite{tsuji08}).  
It is also feasible to observe the CO emission line spectrum expected 
from the MOLsphere, if we carry out aperture-synthesis imaging with 
high-spectral resolution with AMBER, as demonstrated for the red supergiants 
Betelgeuse and Antares (Ohnaka \cite{ohnaka13b}; Ohnaka et al. 
\cite{ohnaka13a}).  

We assumed local thermodynamical equilibrium (LTE) for modeling the 
MOLsphere.  We examined whether this assumption is valid for the 
derived density and temperature of the MOLsphere.  
As described in Ohnaka (\cite{ohnaka04}), we compared the collisional and 
radiative de-excitation rates.  The collisional de-excitation rate 
$C_{\rm ul}$ is estimated as $N \sigma_{\rm ul} \varv_{\rm rel}$, where $N$, 
$\sigma_{\rm ul}$, and $\varv_{\rm rel}$ denote the number density of the 
primary collision partner, collisional cross section, and relative 
velocity between the collision partner and CO molecules.  
We approximate $\sigma_{\rm ul}$ with the geometrical cross section 
of $10^{-15}$~cm$^2$ and adopt $\varv_{\rm rel}$ = 5~\KMS.  
The number density of the primary collision parter (H$_2$ or H) is 
estimated as follows.  
We estimate the CO number density by dividing the derived CO column density 
by the geometrical thickness of the MOLsphere.  
However, we arbitrarily set the geometrical thickness to 0.1~\RSTAR\ 
because this parameter cannot be constrained from the current data. 
Therefore, we estimated a lower limit on the CO number density by 
dividing the derived CO column density ($10^{20}$~\PERSQCM) by the geometrical 
distance between the stellar surface and the radius of the MOLsphere.  
This results in a CO number density of $2.2\times10^{7}$~\PERCBCM.  
In chemical equilibrium at 1500~K, this CO number density suggests 
H$_2$ and H number densities of $5.6\times10^{10}$ and 
$9.5\times10^{9}$~\PERCBCM, respectively.  
Using the H$_2$ number density for $N$, we obtain 
$C_{\rm ul} = 2.8 \times 10$~s$^{-1}$.  This collisional de-excitation rate 
is much higher than the radiative de-excitation rates $A_{\rm ul}$ = 
0.6--1~s$^{-1}$.  Therefore, it is unlikely that our modeling is significantly 
affected by NLTE effects.

\begin{figure}
\resizebox{\hsize}{!}{\rotatebox{0}{\includegraphics{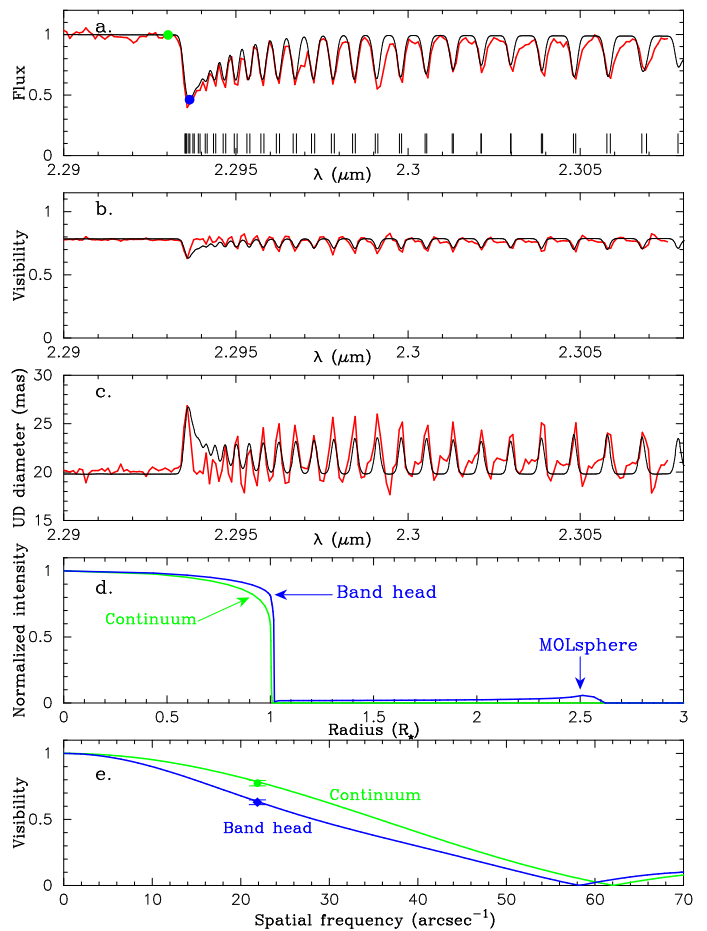}}}
\caption{
Comparison of the MARCS+MOLsphere model with the AMBER data of Aldebaran 
shown in the same manner as in Fig.~\ref{marcs_model}. 
}
\label{wme_model}
\end{figure}

\begin{figure}
\resizebox{\hsize}{!}{\rotatebox{0}{\includegraphics{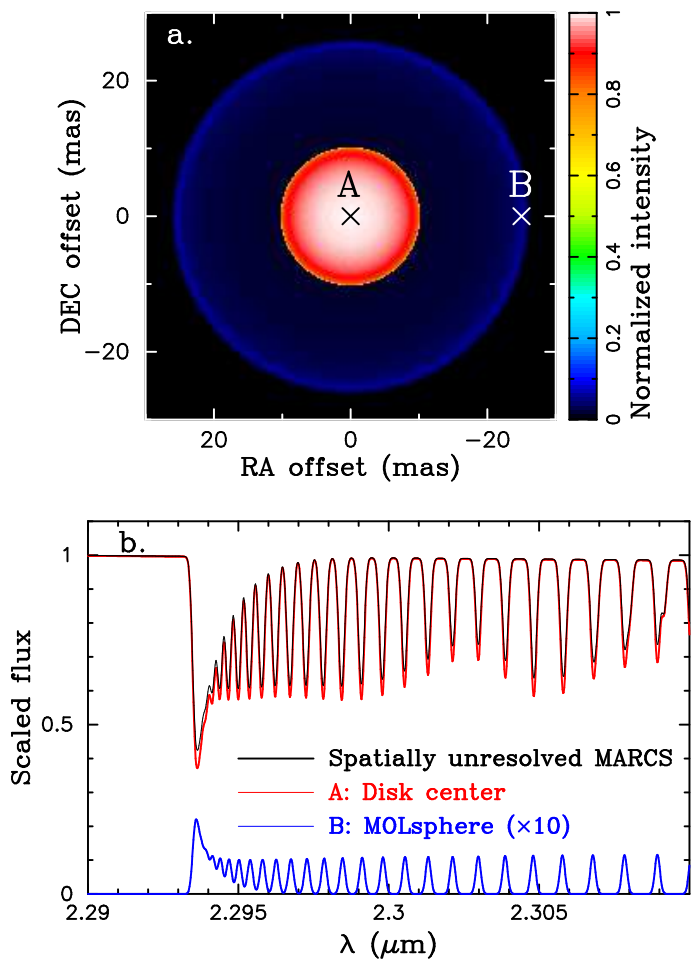}}}
\caption{
{\bf a:} The image predicted at the CO band head by the MARCS+MOLsphere 
model shown in Fig.~\ref{wme_model}.  
{\bf b:} Spatially resolved spectra at the center of the stellar disk 
(position A in the panel {\bf a}) and at the edge of the MOLsphere (position 
B) are shown by the red and blue solid line, respectively.  
The spatially unresolved spectrum from the MARCS-only model 
is plotted by the black solid line. 
The spectrum at the edge of the MOLsphere is scaled up by a factor of 10.  
}
\label{alftau_1dspec}
\end{figure}

\section{Discussion}
\label{sect_discuss}


Our modeling of the spatially and spectrally resolved CO line data suggests 
a high CO column density in the MOLsphere of Aldebaran.  
Tsuji (\cite{tsuji01}) estimated an \HOH\ column density of 
$2\times10^{17}$~\PERSQCM\ from the weak absorption feature near 6.6~\micron. 
The CO number density estimated above suggests an \HOH\ column density of 
$1.4\times10^{20}$~\PERSQCM, if we assume chemical equilibrium at 1500~K.  
The \HOH\ column density implied from our modeling of the CO lines is much higher 
than estimated from the previous 
spectroscopic study.  A possible reason for this disagreement may be that it is 
not straightforward to disentangle the effects of the geometrical extension
and column density from the observed spectra alone.  


Based on the spectral analysis of the CO fluorescent lines in the UV, 
McMurry \& Jordan (\cite{mcmurry00}) deduced the inhomogeneous structure 
of the atmosphere of Aldebaran: the chromosphere as hot as $10^4$~K 
extending into the transition region with $10^5$~K and the cool CO gas at 
$\sim$2000~K. 
This temperature of the cool CO gas is comparable to the temperature 
of the MOLsphere derived above.  
However, the CO column density derived by McMurry \& Jordan (\cite{mcmurry00}) 
is $\sim \! 1 \times 10^{17}$~\PERSQCM, which is much lower than 
derived for the MOLsphere.  
We computed MARCS+MOLsphere models with the CO column density fixed to 
$1\times10^{17}$~\PERSQCM, but the angular size in the CO lines predicted by 
these models is too small compared to 
the observed data, simply because the density is too low.  
Therefore, this discrepancy in the CO column density cannot be reconciled 
by adjusting the parameters in our modeling of the MOLsphere.  
Instead it may be related to the location where the CO fluorescent lines 
originate.  
The CO fluorescent lines in the UV are pumped by the \ion{O}{i} lines that 
form in the chromosphere at temperatures of 6500--8000~K (McMurry 
\cite{mcmurry99}; McMurry \& Jordan \cite{mcmurry00}).  
The spectral analysis of the latter authors implies that ``the \ion{O}{i} 
radiation and CO molecules must be in close proximity''.  
Therefore, the CO column density derived from the fluorescent lines may 
refer only to the cool CO gas near the hot chromospheric gas, while the 
CO column density estimated from the CO first overtone lines refers to 
the entire cool CO gas.  

\begin{figure}
\resizebox{\hsize}{!}{\rotatebox{0}{\includegraphics{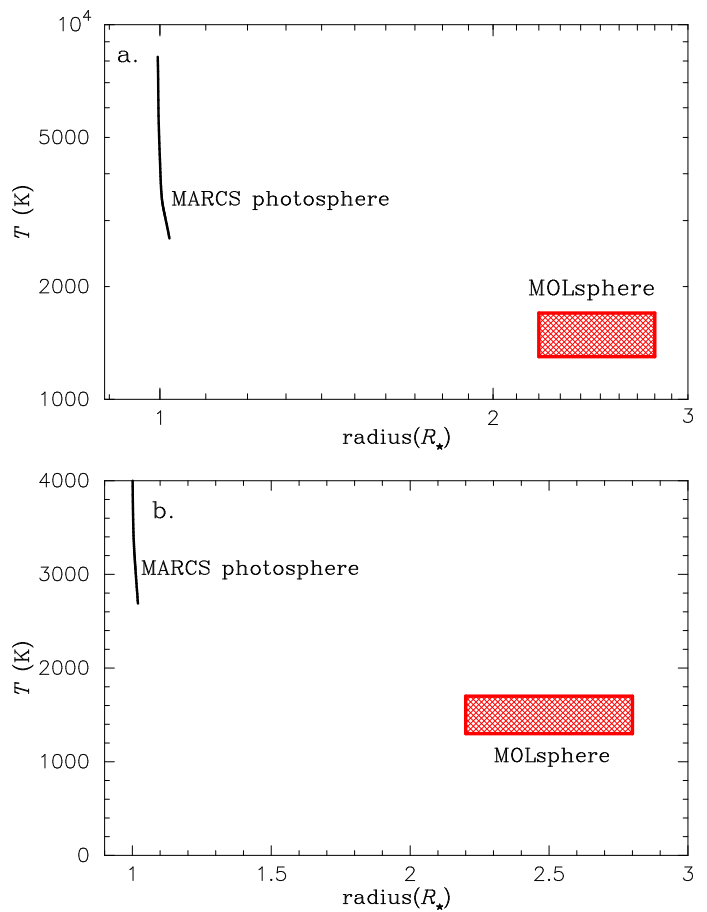}}}
\caption{
Temperature distribution as a function of radius predicted by the 
MARCS model for Aldebaran (black solid line) and the temperature of 
the MOLsphere derived from our modeling (red hatched rectangle).  
The panel {\bf a} shows the entire temperature range covered by the 
MARCS photosphere (note the logarithmic scale both for the temperature 
and radius), 
while the panel {\bf b} shows a lower temperature region including the 
upper photosphere and MOLsphere on the linear scale.  
}
\label{Tplot}
\end{figure}

Figure~\ref{Tplot} shows the temperature distribution of the MARCS 
model for Aldebaran and the temperature and location of the MOLsphere 
derived from our modeling.  Compared to the steep temperature decrease 
in the photosphere, the temperature gradient in the outer atmosphere 
is much smaller.  This is similar to what was found in the M7 giant 
BK~Vir, in which the temperature 
in the MOLsphere is equal to or even higher than in the uppermost layer 
of the MARCS model (Ohnaka et al. \cite{ohnaka12}).  
The CO column density in the MOLsphere of Aldebaran is comparable to the 
value 
found in the red supergiants Betelgeuse (Ohnaka et al. \cite{ohnaka09}; 
\cite{ohnaka11}) and Antares (Ohnaka et al. \cite{ohnaka13a}), as well as 
BK~Vir (Ohnaka et al. \cite{ohnaka12}), 
despite the higher effective temperature and higher surface gravity of 
Aldebaran.  
This implies that the physical process responsible for the formation of 
the MOLsphere may be the same for red supergiants and (non-Mira) red 
giants regardless the spectral type and luminosity class.  

To shed new light on the origin of the MOLsphere, 
the magnetohydrodynamical (MHD) simulation of Suzuki (\cite{suzuki07}) 
is interesting for the present study.  
In particular, his model IV has stellar parameters similar to Aldebaran: 
\TEFF\ = 3900~K, \MSTAR\ = 1~\MSOL, \RSTAR\ = 31~\RSOL, and \LOGG\ = 1.4. 
He simulates Alfv\'en waves, which are excited by the surface convection and 
travel outward along an open 1-D magnetic flux tube from the photosphere to 
$\sim$25~\RSTAR.  The simulation also solves the thermal structure of 
the wind, instead of assuming a given temperature profile.  
The magnetic field used in the simulation 
corresponds to a surface-averaged magnetic field strength of 1~G.  
The stellar wind of the model IV shows hot gas bubbles with a temperature 
higher than $\sim \!\! 10^{5}$~K embedded in much cooler gas.  
A snapshot of the temperature structure (see Fig.~9 in his paper) shows that 
the temperature in the cool gas can be as low as 1000--2000~K, which agrees 
with the observationally derived temperature of the MOLsphere.  
The model also show hot gas bubbles with temperatures of $10^4$--$10^5$~K, 
which may correspond to the chromosphere or the ``buried'' corona 
as Ayres et al. (\cite{ayres03}) propose.  
The lower limit on the CO number density of $2.2\times10^7$~\PERCBCM\ 
derived above suggests a gas density of $1.9 \times 10^{-13}$~\GRPERCBCM. 
The time-averaged gas density predicted at 2~\RSTAR\ for the model IV 
is $\sim \!\! 10^{-14}$~\GRPERCBCM.  Given the large temporal fluctuation 
in the density structure predicted by the model (by a factor of 
10--100) and the uncertainty in the observationally derived CO column density, 
this value roughly agrees with the observationally estimated density. 

However, the time-averaged mass-loss rate predicted by this model is 
$\sim \!\! 10^{-9}$~\MSOLPERYR, which is significantly higher than the 
observationally estimated value of (1--1.6)$\times 10^{-11}$~\MSOLPERYR\ 
(Robinson et al. \cite{robinson98}; Wood et al. \cite{wood07}), 
although the model is not constructed particularly for Aldebaran.   
Also, as pointed out by Airapetian et al. (\cite{airapetian10}), 
the fully ionized plasma is assumed, while the winds from red giants are 
weakly ionized.  
%
%
While the MHD simulation of the stellar wind of Aldebaran by Airapetian et 
al. (\cite{airapetian10}) assumes weakly ionized plasma, 
the temperature structure is not self-consistently computed in their 
simulation, and 
therefore, their model cannot yet account for the MOLsphere component.  
Another uncertainty in the MHD models is that 
the strength of the magnetic fields in Aldebaran is not observationally 
determined.  
On the one hand, the detection of \ion{O}{vi} emission lines 
suggests that the outer atmosphere of Aldebaran is influenced by 
magnetic fields (Harper et al. \cite{harper11}).  
On the other hand, 
the spectropolarimetric observations of Auri\`{e}re et al. 
(\cite{auriere10}) detected no signature of magnetic fields in Aldebaran, 
implying that the surface-averaged field strength is much smaller than 
the 1~G in the model.  
An extension of the MHD simulations of Suzuki (\cite{suzuki07}) with 
weaker magnetic fields and weakly ionized plasma, as well as the 
inclusion of the MOLsphere component in the model of Airapetian et al. 
(\cite{airapetian10}) would be crucial for comparison with the observed 
data.  

McMurry \& Jordan (\cite{mcmurry00}) propose an alternative scenario, 
in which the hot chromospheric gas is associated to acoustic shocks.  
The one-dimensional simulation of acoustic waves for a K5 giant (corresponding 
to Aldebaran) by Buchholz et al. (\cite{buchholz98}) shows that while the 
temperature behind shocks can become higher than $\sim$5000~K, the gas 
between shocks can be as cool as 2000~K.  
However, the cooling due to 
molecules (most importantly CO) is not included in their simulation.  
A more realistic simulation of acoustic waves  including the 
molecular cooling is necessary to understand the formation mechanism of the 
chromosphere and MOLsphere.  

While our MARCS+MOLsphere model simultaneously explains the spectrum and 
angular size in the CO lines of Aldebaran, 
there is still a problem with the MOLsphere 
as mentioned by Ohnaka et al. (\cite{ohnaka12}) for the M7 giant BK~Vir. 
A large number of other molecular species is expected from the MOLsphere, 
most notably TiO.  The contribution of TiO in the MOLsphere can make the 
TiO absorption bands too strong compared to the observed spectra.  
Ohnaka et al. (\cite{ohnaka12}) suggest that the TiO bands may form 
by scattering in the MOLsphere, and the scattered emission may fill in 
the absorption caused by the MOLsphere.  However, it is necessary to 
construct a MOLsphere model with the scattering taken into account and 
compare it with the observed spectra and the available and future angular 
diameter measurements in the TiO bands.  


\section{Conclusions}
\label{sect_concl}

We have spatially resolved the K5 giant Aldebaran in the individual CO 
first overtone lines near 2.3~\micron, taking advantage of high spatial 
and high spectral resolution of VLTI/AMBER.  
The observed uniform-disk diameter in the CO lines is 20--35\% larger 
than in the continuum. 
This increase in the angular size in the CO features would be difficult to 
detect with a spectral resolution lower than $\sim$1500 (AMBER's 
medium spectral resolution mode).  
Our observation of Aldebaran illustrates the power of the high-spectral 
resolution capability of AMBER.  
While the MARCS photospheric model can reproduce the observed CO line 
spectrum very well, it fails to explain the observed angular diameter 
in the CO lines: the model predicts the angular diameter to be too small 
compared to the observation.  
This reveals the presence of the MOLsphere in a K giant for the first time 
in the individual CO lines.  
Our modeling of the AMBER data suggests that the MOLsphere extends to 
$\sim$2.5~\RSTAR\ with a temperature of $\sim$1500~K and 
a CO column density of $\sim \!\! 1\times10^{20}$~\PERSQCM.  
The temperature gradient in the MOLsphere is much smaller than in 
the photosphere.  
The derived temperature and density of the MOLsphere roughly agree 
with the MHD simulation of Suzuki (\cite{suzuki07}), although a 
modeling with more appropriate assumptions for Aldebaran is necessary for 
understanding the origin of the MOLsphere.  
%

Our work demonstrates that the high spatial and high spectral resolution of 
VLTI/AMBER enables us to extract unique information on the outer atmosphere 
even from data on only one baseline.  
Now it is crucial to carry out a systematic survey of red (super)giants 
with a range of effective temperature and luminosity with this technique 
for obtaining a more comprehensive picture 
of the dependence (or absence of it) of the properties of the MOLsphere 
on stellar parameters.  
This is possible with the current AMBER's sensitivity.

\begin{acknowledgement}
The author thanks the ESO VLTI team for supporting our AMBER observation 
and Karl-Heinz Hofmann for the fruitful discussion about the 
reduction of the AMBER data. 
The author also thanks the anonymous referee for his/her constructive 
comments.  
\end{acknowledgement}

\appendix

\section{Computation of the intensity profile, spectrum, and visibility 
from the MARCS+MOLsphere model}
\label{appendix_model}

\begin{figure}
\resizebox{\hsize}{!}{\rotatebox{0}{\includegraphics{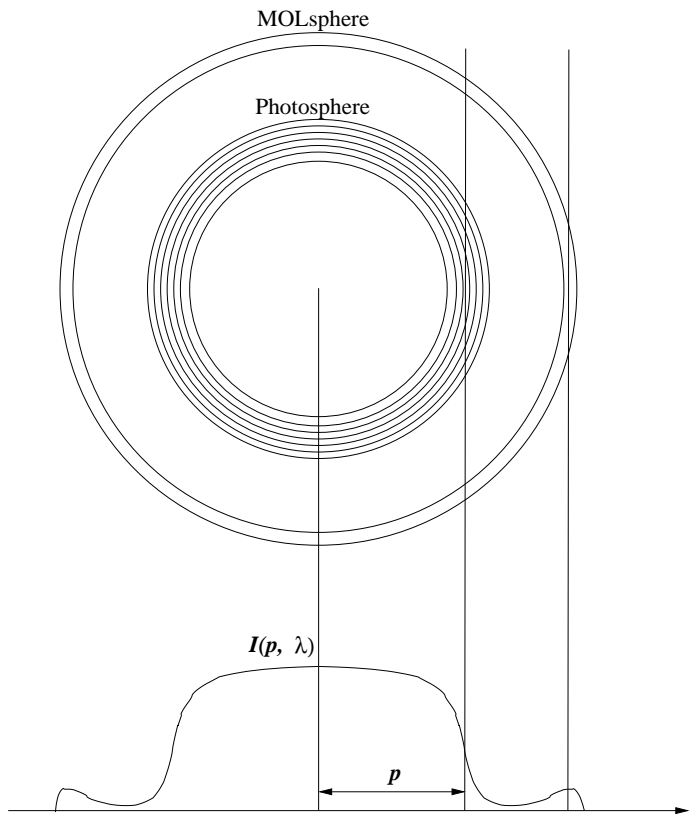}}}
\caption{
Schematic view of the computation of the intensity profile from the 
MARCS+MOLsphere model.  
Each MARCS model consists of 56 layers, but only six layers are drawn 
for the sake of clarity.  
}
\label{model_schematic}
\end{figure}

We first compute the monochromatic intensity profile 
only from a given MARCS model 
with a wavelength interval of $2.7\times10^{-6}$~\micron.  
This wavelength interval was chosen to sample at least 4--5 points 
across the line profile whose width is 
determined by the thermal and micro-turbulent velocity.  
As depicted in Fig.~\ref{model_schematic}, the monochromatic intensity 
$I_{\rm MARCS} (p,\lambda)$ at the wavelength $\lambda$ and at a given 
impact parameter ($p$) is obtained by 
\[
I_{\rm MARCS}(p,\lambda) = \int S_{\lambda}(\tau_{\lambda}) \, e^{-\tau_{\lambda}} \,
d\tau_{\lambda}, 
\]
where $S_{\lambda}$ and $\tau_{\lambda}$ denote the source function and 
the optical depth at $\lambda$ along the ray with the impact 
parameter $p$, and the integration is carried out along this ray.  
In local thermodynamical equilibrium (LTE), the source function is 
given by 
\[
S_{\lambda} = 
\frac{\kappa_{\lambda}}{\kappa_{\lambda}+\sigma_{\lambda}} B_{\lambda} 
+ \frac{\sigma_{\lambda}}{\kappa_{\lambda}+\sigma_{\lambda}} J_{\lambda}, 
\]
where $\kappa_{\lambda}$ and $\sigma_{\lambda}$ represent 
the absorption and scattering coefficients, 
respectively.  $B_{\lambda}$ and  $J_{\lambda}$ represent the Planck function 
and the mean intensity, respectively.  
From the pressure and temperature distributions of a given MARCS model, 
we compute the number of each molecular and atomic species in chemical 
equilibrium.   With the Voigt function adopted for the line profile, 
the absorption coefficient at each wavelength is calculated from the 
CO line list of Goorvitch (\cite{goorvitch94}) and the continuous 
opacities due to the bound-free and free-free transitions of H$^{-}$ and H, 
the free-free transitions H$_2^{-1}$ and He$^{-1}$, and the bound-free 
transitions of Si, Mg, and Ca.  
We used the data of John (\cite{john88}) for the H$^{-1}$ opacity, 
while we adopted the cross sections given in Tsuji (\cite{tsuji71} and 
references therein) for the other continuous opacities.  

The intensity profile from a MARCS+MOLsphere model 
$I_{\rm MARCS+MOL}(p,\lambda)$ is computed using 
the MARCS-only intensity profile $I_{\rm MARCS}(p,\lambda)$ as follows: 
\[
I_{\rm MARCS+MOL}(p,\lambda) = I_{\rm MARCS}(p,\lambda) \, 
e^{-\tau_{\lambda}^{\rm MOL}} 
+ B_{\lambda}(T_{\rm MOL}) (1 - e^{-\tau_{\lambda}^{\rm MOL}}), 
\]
where $\tau_{\lambda}^{\rm MOL}$ denotes the optical depth in the MOLsphere 
along the ray with the impact parameter $p$, and $T_{\rm MOL}$ 
is the temperature of the MOLsphere.  

Once the intensity profile at each wavelength is obtained, the spectrum 
from the MARCS+MOLsphere model ($F_{\lambda}$) is calculated by
\[
F_{\lambda} = 2\pi \int_{0}^{1} I_{\rm MARCS+MOL}(p,\lambda) \, \mu \, d\mu, 
\]
where $\mu$ is defined as $\sqrt{1-(p/R_{\rm MOL})^2}$ with $R_{\rm MOL}$ 
being the outer radius of the MOLsphere.  
The visibility is obtained by taking the Hankel transform (2-D Fourier 
transform for axisymmetric objects) of the intensity profile.  

Finally, the monochromatic intensity profile and visibility, as well as the 
spectrum are spectrally convolved with the AMBER's spectral resolution.  

\end{document}